\documentclass[referee]{aa}
\usepackage{natbib,graphicx,longtable}

\def\Mpc{\hbox{$\rm\thinspace Mpc$}}

\def\hMpc{\hbox{$\thinspace h^{-1}\Mpc$}}
\def\kms{\hbox{$\km\s^{-1}\,$}}

\def\mo{\hbox{${\rm\thinspace M}_{\odot}$}}
\def\etal{{et al.\ }}

\def\kms{\ {\rm km\,s^{-1}}}
\begin{document}
\title{Structure and dynamics of the Shapley Supercluster\thanks{Based on
observations made at the European Southern Observatory, La Silla, Chile,
at the Las Campanas Observatory, Chile, and at the Anglo-Australian
Observatory, Australia.}}
\subtitle{Velocity catalogue, general morphology and mass.}
\author{Dominique Proust\inst{1}
\and Hern\'an Quintana\inst{2} \and Eleazar R. Carrasco \inst{3}
\and Andreas Reisenegger\inst{2} \and Eric Slezak\inst{4} \and
Hern\'an Muriel\inst{5} \and Rolando D\"unner\inst{2} \and Laerte
Sodr\'e Jr.\inst{6} \and Michael J. Drinkwater\inst{7} \and Quentin
A. Parker\inst{8} \and Cinthia J. Ragone\inst{5}}

\institute {GEPI, Observatoire de Paris-Meudon F92195 Meudon CEDEX,
France. \and Departamento de Astronom\'ia y Astrofisica, Pontificia
Universidad Cat\'olica de Chile, Casilla 306, Santiago 22, Chile.
\and Gemini Observatory, Southern Operations Center c/o AURA,
Casilla 603, La Serena, Chile. \and Observatoire de Nice, 06304 Nice
CEDEX4, France. \and Grupo de Investigaciones en Astronom{\'\i}a
Te\'orica y Experimental, Observatorio Astr\'onomico, Laprida 854,
5000 C\'ordoba, Argentina, and CONICET, Buenos Aires, Argentina.
\and Instituto de Astronomia, Geof\'isica e Ci\^encias
Atmosf\'ericas, R. do Mat\~ao 1226, CEP 05508-090 S\~ao Paulo/SP
Brazil. \and Department of Physics, University of Queensland, QLD
4072, Australia, \and Department of Physics, Macquarie University,
NSW 2109, Australia, and Anglo-Australian Observatory, PO Box 296,
Epping NSW 1710, Australia.}

\offprints{Dominique Proust, \email{dominique.proust@obspm.fr}}
\date{Received /Accepted}

\abstract{We present results of our wide-field redshift survey of
galaxies in a 285 square degree region of the Shapley Supercluster
(SSC), based on a set of 10529 velocity measurements (including 1201
new ones) on 8632 galaxies obtained from various telescopes and from
the literature. Our data reveal that the main plane of the SSC
($v\approx 14500\kms$) extends further than previous estimates,
filling the whole extent of our survey region of 12~degrees by
30~degrees on the sky ($30\times 75~h^{-1}{\rm Mpc}$). There is also
a connecting structure associated with the slightly nearer
Abell~3571 cluster complex ($v\approx 12000\kms$). These galaxies
seem to link two previously identified sheets of galaxies and
establish a connection with a third one at $\overline v= 15000\kms$
near R.A.$= 13^{\rm h}$. They also tend to fill the gap of galaxies
between the foreground Hydra-Centaurus region and the more distant
SSC. In the velocity range of the Shapley Supercluster ($9000\kms <
cz < 18000\kms$), we found redshift-space overdensities with
$b_j<17.5$ of $\simeq~5.4$ over the 225~square degree central region
and $\simeq~3.8$ in a 192~square degree region excluding rich
clusters. Over the large region of our survey, we find that the
intercluster galaxies make up 48 per cent of the observed galaxies
in the SSC region and, accounting for the different completeness,
may contribute nearly twice as much mass as the cluster galaxies. In
this paper, we discuss the completeness of the velocity catalogue,
the morphology of the supercluster, the global overdensity, and some
properties of the individual galaxy clusters in the Supercluster.
{\keywords: Galaxies: clusters: individual: Shapley Supercluster,
A3558 --- Galaxies: distances and redshifts --- Cosmology:
observations --- Large-scale structure of the Universe}}
\titlerunning{Shapley Supercluster}
\authorrunning{D. Proust et al.}
\maketitle
\section{Introduction}
In the past few decades, large galaxy redshift surveys have revealed
structures on ever-increasing scales. The largest coherent
structures found are superclusters, collections of thousands of
galaxies with linear sizes as large as 100~Mpc. They offer us
information about the late evolution of the Universe and the
transition from the linear to the non-linear regime. Detailed
investigations have shown the reality of such physical systems with
gravitationally assembling clusters. Galaxy distributions and weak
lensing showed that the mass distribution in superclusters is in
good agreement with the distribution of early-type galaxies in these
structures (Einasto \etal 2003a; Oguri \etal 2004). The mere
existence of these structures places important constraints on
theories of the formation of galaxies and clusters. The Shapley
Supercluster, the subject of this paper, is one of the most massive
concentrations of galaxies in the local Universe (Scaramella \etal
1989; Raychaudhury 1989). It therefore deserves a special
examination of its organisation and it is also of particular
interest to consider its effect on the dynamics of the Local Group.

The Shapley Supercluster (SSC) is a remarkably rich concentration of
galaxies centred around R.A.$=13^{\rm h}25^{\rm m}$, Dec $=
-30^{\circ}$, which has been investigated by numerous authors since
its discovery in 1930 (see Quintana \etal 1995, 2000). It consists
of many clusters and groups of galaxies in the redshift range $0.04
< z < 0.055$. The SSC lies in the general direction of the dipole
anisotropy of the Cosmic Microwave Background (CMB; Smoot \etal
1992), and is located $\sim 100$\hMpc\ beyond the Hydra-Centaurus
supercluster, which in turn is $\simeq 40$\hMpc\ away from us (with
the Hubble constant parameterized as $H_0=100~h\kms~{\rm
Mpc}^{-1}$). Quintana \etal (1995) estimated that (for $\Omega_{\rm
M}= 0.3$, $\Omega_\Lambda=0$, and $H_0= 75\kms {\rm Mpc}^{-1}$) the
gravitational pull of the supercluster may account for up to 25\% of
the peculiar velocity of the Local Group required to explain the CMB
dipole anisotropy, in which case the mass of the supercluster would
be dominated by intercluster dark matter. A major study of the SSC
core region was made by Bardelli \etal (2000, 2001 and references
therein). They derived a total mean overdensity of the SSC of
$N/\overline N \sim 11.3$ on a scale of $10.1h^{-1}$ Mpc and found
that the central part of the supercluster contributes about 26$\kms$
to the peculiar velocity of the Local Group, i.e., 7\% of the total
tidal bulk velocity of $366 \pm 125\kms$ from Hoffman \etal (2001).
The central cluster A3558 has also been observed with the ROSAT PSPC
by Bardelli \etal (1996), who derive a total mass of $M_{\rm
tot}=3.1\times10^{14}~h^{-1}\mo$ within a radius of 2\hMpc. Several
other luminous X-ray clusters are part of the Shapley Supercluster
(Pierre \etal 1994).

The early studies of the Shapley Supercluster were limited
(primarily by observational constraints) to the various rich Abell
galaxy clusters in the region. This approach might give a very
biased view of the overall supercluster, as these clusters represent
only the most concentrated peaks in the luminous matter
distribution. An analysis based on 3000~galaxy redshifts in an area
of about $12\deg \times 8\deg$ (Quintana \etal 2000) was published
by Reisenegger \etal (2000). They estimated an upper bound on the
mass of the central region (within a radius of $8h^{-1}{\rm Mpc}$)
and found the overdensity to be substantial, but insufficient to
contribute more than a small fraction ($3\Omega_m^{-0.4}$\%) of the
observed motion of the Local Group.

In this paper, we make an investigation of the large-scale
distribution of galaxies within the entire SSC region and nearby
regions, using data from long-slit and wide-field multi-fiber
spectrographs. Most of these observational data have already been
published. A large survey was carried out over several years with S.
Shectman's fiber spectrograph mounted on the 100'' DuPont telescope
at Las Campanas Observatory (LCO). Quintana \etal (2000) reported
2868 new velocities measurements, corresponding to 2627~different
galaxies observed at LCO. The complete LCO data, which represent
40\% of the catalogue, is being published (Quintana \etal 2005),
including a new (and last) set of 1201 velocities on 1166 galaxies
observed with the same instrument between 1997 and 1999. Another
fraction of redshifts have come from the FLAIR facility and the UKST
at the Anglo Australian Observatory (AAO). With such multiplexed
facilities, we were able to measure more galaxy redshifts over large
angular extents and obtain a more complete picture of the
composition and distribution of galaxies in the entire supercluster.
Combined with already published redshift sets from several surveys
and papers (Quintana \etal 1995, 1997, 2000; Bardelli \etal 1996,
1998, 2001; Drinkwater \etal 1998, 1999, 2004; the FLASH survey of
Kaldare \etal 2003, and the 6dF survey of Jones \etal 2004), we
built up the most complete velocity catalogue for the SSC,
containing 10529~velocity measurements for 8632~galaxies.

The complete galaxy velocity catalogue is described in Section~2.
This section also includes comparisons between the galaxies in
common among different data sets, as well as the velocity zero-point
shifts required to end up with a homogeneous catalogue at our
disposal. In Section~3, we discuss the completeness of this
catalogue and analyse the galaxy number density over the whole and
the intercluster survey regions. In Section~4 we analyse the
three-dimensional morphology of the SSC. In Section~5, we discuss
properties of the individual Abell clusters, and in Section~6 we
determine the global luminosity and mass of the SSC and its
contribution to the peculiar motion of the Local Group with
respect to the cosmic microwave background. 

\section{The velocity catalogue}

We investigated the large-scale distribution of galaxies throughout
the entire SSC region as defined in Quintana \etal (2000) and close
environs using data from long-slit and wide-field multi-fibre
spectrographs at various telescopes. Most of these data are related
to 3 major surveys: 40\% from Las Campanas observations (Quintana
\etal 2005), 24\% from the UKST at Siding-Spring, Australia
(Drinkwater \etal 1999, 2004, Kaldare \etal 2003, Jones \etal 2004),
15.5\% from the ESO 3.6~m telescope at La Silla, Chile (Bardelli
\etal 1997, 1999, 2001; Quintana \etal 1997). We also searched the
literature for velocities between $11^{\rm h}30^{\rm m} <$ R.A. $<
14^{\rm h}30^{\rm m}$ and $-23^\circ >$ Dec $> -45^\circ$. A total
of 1520 galaxies have been observed more than once, which allows us
both to obtain more accurate velocity determinations for these
galaxies and to compare different data sets for assessment of the
general quality of the data and shifts in the velocity zero-point.
Combined with already published redshift sets from several surveys
and papers, we obtain the most complete velocity catalogue for the
SSC area, containing 10529~velocities for 8632~galaxies, which are
now available from the NED database\footnote{The NASA/IPAC
Extragalactic Database (NED) is operated by the Jet Propulsion
Laboratory, California Institute of Technology, under contract with
the National Aeronautics and Space Administration}. The main point
before any analysis is to combine these different data sets to build
a homogeneous catalogue. Below, we comment on the detailed sources
and instruments used in this catalogue.

\subsection{Las Campanas data.}

The spectroscopic observations were carried out using the fiber spectrograph
mounted on the 2.5m (100'') du Pont telescope at Las Campanas
Observatory\footnote{The Las Campanas Observatory is operated by the Carnegie
Institution of Washington} in several observing sessions ranging from 1992 to
1999. Some of them are already published in Quintana \etal (1995, 1997).
2868~new measurements are described and compared with other sources in
Quintana \etal (2000). A set of 1201~new velocities for 1166 different
galaxies has been obtained between 1997 and 1999. All these data are described
and published in Quintana \etal (2005).

\subsection{UKST/FLAIR-II, 6dF and FLASH data.}

Two sets of published data were obtained with the UKST/FLAIR-II
system at the Anglo Australian Observatory (Drinkwater \etal 1999,
2004). They correspond to 710~galaxies observed over 7~UKST standard
fields in the SSC region, covering an area of 182 deg$^{2}$. The
target galaxies were originally obtained from the red ESO/SRC sky
survey plates scanned by the Paris Observatory MAMA plate-measuring
machine. The galaxy sample was defined to a photometric limit of $R=
16$, corresponding (assuming a mean $B-R=1.5$) to $B= 17.5\approx
b_j$, the nominal galaxy limiting magnitude of the FLAIR-II system
(Parker \& Watson 1995). This corresponds to an absolute magnitude
of $M_B=-18.3+5\log h$ at the Shapley distance of $145h^{-1}\Mpc$.
The data were reduced using the dofibers package in IRAF (Tody
1993). Redshifts were measured for galaxy spectra using the
cross-correlation task XCSAO in RVSAO (Kurtz \& Mink 1998) using a
mixture of a dozen stellar and galaxy templates.

The Six-Degree Field Galaxy Survey (6dFGS) of the Southern sky,
currently in progress, aims to measure the redshifts of around
150000 galaxies, and the peculiar velocities of a 15000 galaxy
sub-sample, over almost the entire southern sky (Jones \etal 2004).
Among this large target sample, 1328 remaining candidates from the
above FLAIR-II program were selected in the SSC area. The 6dF fibre
spectroscopy facility installed at the UKST has up to now observed
584 galaxies in the Shapley region with reliable redshifts. The
reduction of the spectra was made using an adapted version of the
2dF Galaxy Redshift Survey software (Colless \etal 2001).

To complete the set of velocities obtained from the UKST, we have
also included observations from the FLASH (FLAIR Shapley-Hydra)
redshift survey of Kaldare \etal (2003). It consists of 4613
galaxies brighter than $b_{j}$= 16.7 over a 700~square degree region
of sky in the general direction of the Local Group motion. The
survey region is a $70^\circ \times 10^\circ$ strip spanning the sky
from the SSC to the Hydra cluster and contains 3141 galaxies with
measured redshifts. The reduction procedure was similar to that used
above for FLAIR-II spectra (Drinkwater \etal 1999, 2004).

After eliminating already published velocities from NED and galaxies
outside the SSC region, we add 1411 new velocities to our database.

\subsection{ESO observations}

A major study of the SSC was made by Bardelli \etal (1994, 1996,
1998, 2000, 2001) and Baldi \etal (2001). All their observations
were made at the ESO, La Silla (Chile) 3.6~m telescope, initially
equipped with the multiobject facility OPTOPUS (Lund 1986), and
later with MEFOS (Felenbok \etal 1997). Their analyses are based on
data from the literature and on 4~sets of new velocities, 311~in
Bardelli \etal (1994), 174~in Bardelli \etal (1998), 442~in Bardelli
\etal (2000), then 662~for 581~new galaxies in Bardelli \etal
(2001), with a total of 1589~new velocities, representing 15.5\% of
the velocity catalogue. Their sample is homogeneous, in a well
defined magnitude range ($17.0~<~b_{j}~<~18.8$). Their analyses are
mainly concerning the SSC central region based on a sample of
$\simeq$~1300 spectra covering an area of $\simeq~26~{\rm deg}^{2}$.

Another survey centered on 15~Abell cluster targets in the direction
of the SSC has been carried out by our group (Quintana \etal 1997),
also using the multiobject facility MEFOS at the ESO 3.6~m
telescope. An additional set of 179 velocities has been obtained and
is included in our general database.

\subsection{Comparison of the velocities.}

Comparison of the velocities was made using the 1520~galaxies of the
database with more than one velocity measurement. In order to
combine the new data described above with those already available in
the NED database, we normalized all radial velocity sets to a common
zero point as in Quintana \etal (2000). If the zero points of
different instruments differed by significant amounts, we could have
introduced serious systematic errors in the resulting velocity
dispersion and in the dynamical analysis. Table~1 summarizes the
results of the zero-point shifts obtained from comparison of the new
data with the previously published ones, similar to Table~2 in
Quintana \etal (2000). Finally, we have applied the zero-point
shifts given in Table~1 to all velocities of the catalogue, before
combining the measurements available for each galaxy into a final
average value. For references without velocities in common with our
data, or when the number of velocities was to low to obtain a
meaningful shift, we compared the data with other datasets in the
literature, and calculated the zero-point shift in a transitive way.
The average velocity for galaxies with multiple references was
calculated following Quintana \etal (2000). In a few cases, we
discarded a velocity when analysis of the spectra showed
discrepancies which differed by more than $200~\kms$. The final
sample resulting from this procedure contains 8632~galaxies.

\begin{table*}
\caption{Comparison of the velocities and Zero Point shift between new
velocities and data already published.}
\vspace{0.2cm}
\begin{tabular}{lrrrrr}
\hline
References & $N_{ref}$ & $N_{comm}$ & $\Delta v$ & $\sigma_{\Delta v}$ &
rms \\
           &           &            &  $\kms$ &   $\kms$           & $\kms$ \\
\hline
Bardelli \etal (2001)     &  581 &  48  &    -6   &   15      &    98       \\
Kaldare \etal (2003)      & 1411 &  82  &    -1   &   21      &   103       \\
Drinkwater \etal (2004)   &  404 &  17  &    11   &   15      &    61       \\
Jones \etal (2004)        &  584 &  29  &     8   &   13      &    73       \\
Quintana \etal (2005)     & 1201 &  39  &    18   &   23      &    86       \\
\hline
\end{tabular}
\end{table*}

\section{Observed Galaxy Sample, completeness and overdensity.}

For the purpose of our analysis below, we compare the velocity
catalogue to the total magnitude-limited galaxy distribution in the
survey region. We chose to use the new SuperCOSMOS sky surveys
(Hambly et al.\ 2001a,b,c) to construct the parent photometric
galaxy catalogue for the whole region covered by our nine fields
observed with UKST. We could cross-correlate almost each object with
a SuperCOSMOS galaxy within a matching radius of 8~arcsec. The 6\%
remaining galaxies were visually checked on the sky plates. In this
paper, we quote and use in the analyses the SuperCOSMOS $b_{j}$
magnitudes of all galaxies. We compared the magnitude distributions
of the objects identified in the nine UKST fields, 381 to 384 and
442 to 446: two-sample K-S tests showed that they were all
consistent with the same distribution (mean $b_{j} \simeq 17.2$).

Allowing for the boundaries of the survey fields and a circular
region 1~degree in diameter, which we excluded around the bright
star HD 123139 (at R.A.$=14^{\rm h}06^{\rm m}41^{\rm s}$, Dec$=
-36^\circ 22'12''$, J2000), the total survey region has an area of
225 square degrees. We also defined a restricted intercluster region
by excluding regions 1~degree in diameter around any rich galaxy
clusters in the Shapley velocity range, leading to an area of 192
square degrees.

\subsection{Completeness of the velocity catalogue.}

In such a compilation, it is evident that different authors had different,
often not easily quantifiable selection criteria. This fact makes the
completeness analysis given here necessary, but, on the other hand, it makes it
impossible to do it perfectly, because the many biases cannot be fully grasped
by the variables RA, Dec, magnitude and cluster
membership.
Thus, some biases (such as the one due to morphology) are
necessarily left even after completeness corrections are done.

We used the parent photometric galaxy catalogue to determine the
completeness of the velocity catalogue as a function of 4 limiting
magnitudes, respectively $b_{j}$ = 17.0, 17.5, 18.0, and 18.5. This
is shown in Table~2 for each UKST plate and in Table~3 for both the
full region and the intercluster region. This table shows that the
completeness is highest for the brighter magnitude limits, peaking
at 61\% for the whole sample at $b_{j}<17.0$. Figure~1 shows the
completeness of the Shapley velocity catalogue. Each of the panels
corresponds to a different magnitude limit: 17, 17.5, 18, and 18.5.
We used the HEALPix package\footnote{http://www.eso.org/science/healpix/content/HEALPix\_reference.html}
(G\'orski et al. 2005) to perform the completeness
mask. For each pixel of $2.6 \times 10^{-4}~strd$, we count galaxies up to a
magnitude limit, defining the completeness in a pixel as the ratio
of the number of galaxies with measured velocities to the number of
SuperCOSMOS galaxies in the same pixel. The scale represents the
fraction of galaxies in the SuperCOSMOS catalogue that have measured
redshifts in the present velocity catalogue.

\begin{figure*}[htb]
\centering
\includegraphics[height=11.0cm]{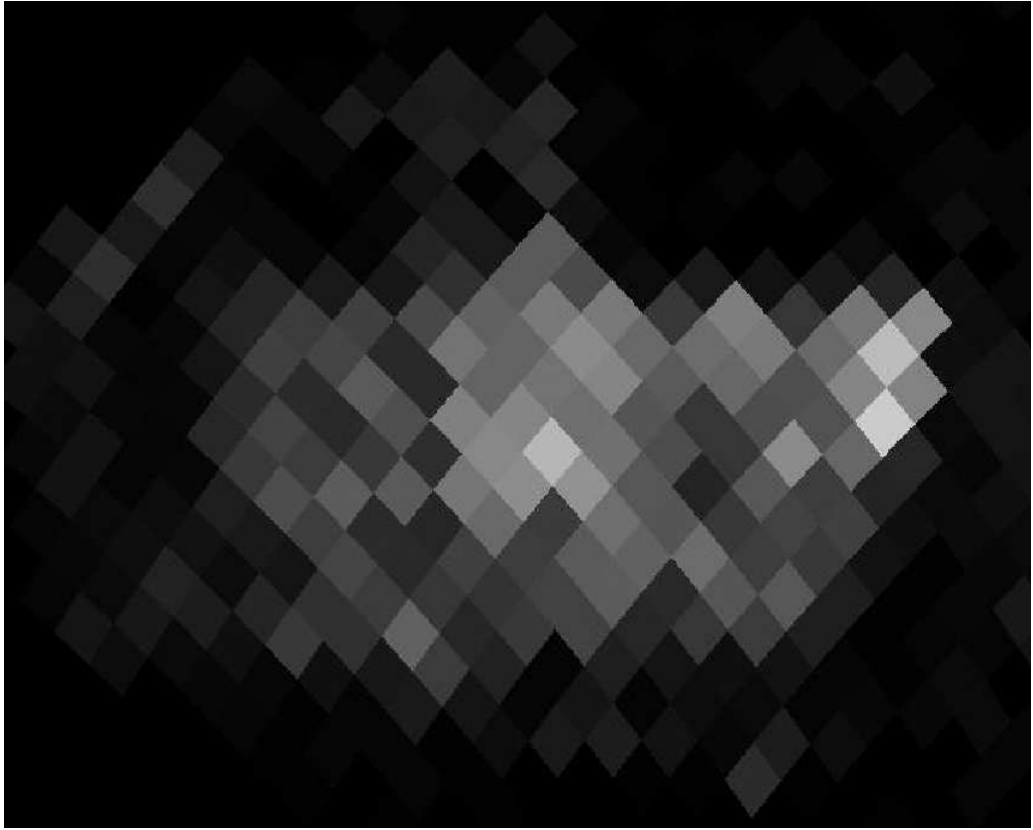}
\caption{Completeness of the Shapley velocity catalogue with
different magnitude limits: $b_{j}$= 17.0 (top left), 17.5 (top
right), 18.0 (bottom left), and 18.5 (bottom right).}
\end{figure*}

\begin{table}
\caption{Completeness for each of the 9 Shapley ESO/UKST fields for
four $b_{j}$ magnitude limits (all galaxies brighter than the given
limit).} \label{tab-obs}
\begin{center}
\vspace{0.2cm}
\begin{tabular}{lrrrrr}
\hline
Name     & {\bf F446} & {\bf F445} & {\bf F444} & {\bf F443} & {\bf F442}  \\
R.A.     &  $14^{\rm h}11^{\rm m}$   &  $13^{\rm h}48^{\rm m}$   &  $13^{\rm h}25^{\rm m}$   &  $13^{\rm h}02^{\rm m}$   &  $12^{\rm h}39^{\rm m}$    \\
Dec      &  $-30^\circ$ &  $-30^\circ$ &  $-30^\circ$ &  $-30^\circ$ &  $-30^\circ$  \\
         &            &            &            &            &             \\
$b_{j}< 17.0$ & 0.29  &    0.67    &   0.92     &    0.94    &    0.33     \\
$b_{j}< 17.5$ & 0.17  &    0.48    &   0.79     &    0.79    &    0.20     \\
$b_{j}< 18.0$ & 0.10  &    0.30    &   0.61     &    0.62    &    0.10     \\
$b_{j}< 18.5$ & 0.06  &    0.19    &   0.44     &    0.43    &    0.07     \\
\hline
\end{tabular}
\begin{tabular}{lrrrrr}
\hline
Name     &~~& {\bf F384} & {\bf F383} & {\bf F382} & {\bf F381} \\
R.A.     &~~&   $14^{\rm h}00^{\rm m}$~ &   $13^{\rm h}36^{\rm m}$  &   $13^{\rm h}12^{\rm m}$  &   $12^{\rm h}48^{\rm m}$  \\
Dec      &~~&  $-35^\circ$ &  $-35^\circ$ &  $-35^\circ$ &  $-35^\circ$ \\
         &~~&            &            &            &            \\
$b_{j}< 17.0$ &~~& 0.33  &    0.65    &   0.60     &    0.26    \\
$b_{j}< 17.5$ &~~& 0.19  &    0.51    &   0.48     &    0.16    \\
$b_{j}< 18.0$ &~~& 0.10  &    0.29    &   0.33     &    0.11    \\
$b_{j}< 18.5$ &~~& 0.07  &    0.18    &   0.21     &    0.05    \\
\hline
\end{tabular}
\end{center}
\end{table}
\bigskip

\begin{table}
\caption{Global completeness of the velocity catalogue for 4 $b_{j}$
magnitude limits. The completeness of each sample compared to the
SuperCOSMOS catalogue is given in parentheses (all galaxies brighter
than the given limit) and brackets [galaxies brighter than the given
limit, but fainter than the limit in the previous row].}
\label{tab-samples}
\begin{center}
\begin{tabular}{lrrl}
\hline
Field  & $b_{j}$ Mag lim. &  SuperCOSMOS  &  Velocities \\
\hline
Full          & $< 17.0$ &  2981 &  2332 (78\%)  \\
Full          & $< 17.5$ &  6331 &  3601 (57\%) [38\%] \\
Full          & $< 18.0$ & 13361 &  4843 (36\%) [18\%] \\
Full          & $< 18.5$ & 27177 &  5701 (21\%) [ 6\%] \\
\hline
Inter-cluster & $< 17.0$ &  2404 & 1547  (64\%)  \\
Inter-cluster & $< 17.5$ &  5105 & 2411  (47\%) [32\%] \\
Inter-cluster & $< 18.0$ & 10862 & 2890  (27\%) [ 8\%] \\
Inter-cluster & $< 18.5$ & 21917 & 3233  (15\%) [ 3\%] \\
\hline
\end{tabular}
\end{center}

\end{table}

\subsection{Galaxy Overdensity}

The present velocity catalogue allows us to calculate the
overdensity in redshift space over our full survey region of 225
deg$^2$ and our intercluster region of 192 deg$^2$. As we are mostly
interested in galaxies at the distance of the SSC, the overdensity
is best seen as the peaks in the velocity histogram. We also
computed the expected distribution for a smooth, homogeneous galaxy
distribution based on the number count data of Metcalfe \etal
(1991), allowing for the differential incompleteness of each sample
as a function of $b_j$, as listed in the final column of Table~3.


We calculated the galaxy overdensity as the ratio of the number of
observed galaxies within the nominal velocity limits of the SSC
complex ($9000-18000\kms$) to the number expected from the Metcalfe
counts within the same velocity limits, all with $b_j<17.5$, since,
at the distance of the SSC, this magnitude limit contains
essentially all galaxies that make a substantial contribution to the
total luminosity. In the full region, the overdensity is\footnote{We
give formal, statistical errors, not including any systematic
effects. In particular, these overdensities are likely to be
overestimates, as the galaxies have generally been selected in such
a way as to maximize their chance of belonging to the SSC.} $5.4\pm
0.2$, and in the restricted intercluster region it is $3.8 \pm 0.2$.
The {\it surface} overdensity in the area covered by this study
(disregarding redshift information) is $2.9\pm 0.2$.

This strengthens earlier evidence for a very significant galaxy
overdensity in the intercluster space of the SSC region (e. g.,
Bardelli \etal 2000). In terms of galaxy numbers, the intercluster
galaxies make up 48\% of the 3705 galaxies in the SSC velocity
range. So, assuming a similar mass function and taking account of
the lower completeness in the intercluster regions, these contribute
nearly twice as much mass as the cluster galaxies.

In their study of the central part of the SSC, Bardelli \etal (2000)
report an overdensity of $3.9 \pm 0.4$ for their intercluster sample
and $11.3 \pm 0.4$ for their total sample, on scales of
$10h^{-1}$Mpc. The latter value of the overdensity is much higher
than ours, paper due to the much smaller scale. Drinkwater \etal
(2004) found overdensities respectively $5.0 \pm 0.1$ and $3.3 \pm
0.1$ for a magnitude limit $R<17.0$ in a slightly less extended
region of the SSC, of $26-40h^{-1}$Mpc. These two sets of values are
comparable, since $b_j - R \simeq$~0.7 in the direction of the SSC.
They are higher than the mean density of luminous matter in the
superclusters of the SDSS (Einasto \etal 2003a). The peak densities
are in the range of 2.2-4.5, and the mean densities in the range of
1.9-2.8 as computed by Einasto (2005). Our data probe the
distribution on much larger scales of around 12--18 degrees,
corresponding to $30-46h^{-1}$Mpc, and not just in the denser, inner
region. For the intercluster regions of the Horologium-Reticulum
supercluster (HRS) on a large scale $~40h{-1}$Mpc, Fleenor \etal
(2005) find an overdensity of 2.4. Their data and those from the
present paper support the conclusion of Einasto \etal (2001) that
the SSC and HRS constitute the two largest mass concentrations in
the local universe.

\section{Velocity distribution and topology of the Shapley Supercluster.}

The definition of the topology of the Shapley Supercluster is not an
easy task, because of the complexity of the structures in the
velocity distribution. The presence of many clusters, with their
characteristic finger-of-God velocity structures, makes the study
difficult. Moreover, remaining irregularities and gaps in the
observations could mimic apparent structures. Finally, as modern
redshift surveys show, dense structures are linked to each other by
filaments and walls, forming a fabric that weaves throughout space.

\subsection{Velocity distribution}

Fig.~2 shows the velocity distribution of galaxies in the direction
of the Shapley Supercluster with all available velocities in the
range $0 \leq v \leq 30000~\kms$, with a step size of $500~\kms$.
The histogram presents four maxima, as discussed by Quintana \etal
(2000). The main body of the SSC is represented by the highest peak,
which is centered at $\simeq 15000~\kms$ and extends over the
velocity range $13000 - 18000~\kms$. Another peak between $9000$ and
$13000~\kms$ shows the nearer concentration which is located to the
East, centered on A3571. The other two peaks are at $4000~\kms$ (the
Hydra-Centaurus region), and at $23000~\kms$, another possible
supercluster behind the SSC. At higher redshifts, a few clumps of
galaxies are also present.

\begin{figure*}[htb]
\centering
\includegraphics[width=8.0cm,angle=-90]{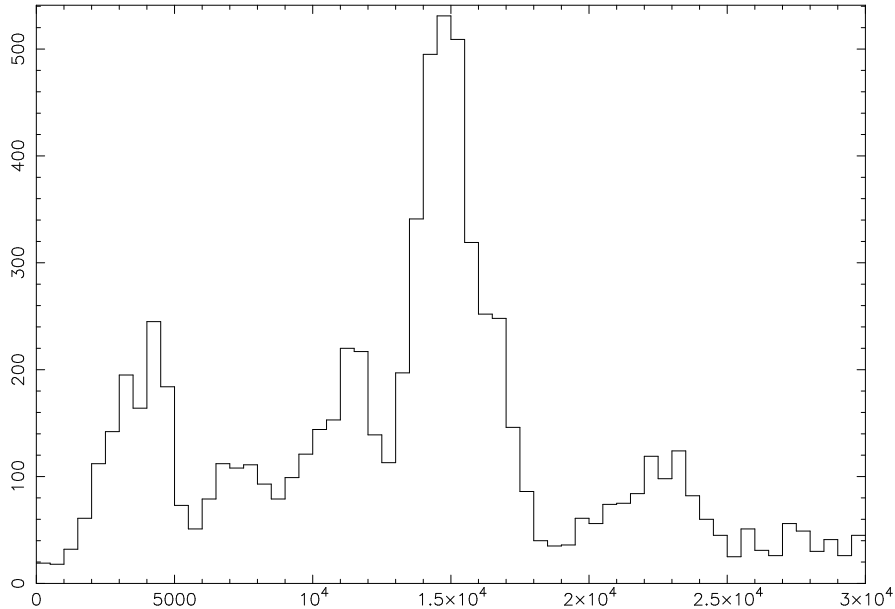}
\caption{Histogram of galaxy velocities in the direction of the Shapley
Supercluster with all velocities available in the range
$0~\kms \leq v \leq 30000~\kms$, with a step size of $500~\kms$.}
\end{figure*}

Figure~3 shows the combined resulting distribution of galaxies
towards the Shapley Supercluster as wedge diagrams in right
ascension (top) and declination (bottom) for the whole velocity
catalogue. The importance of the SSC in this region of the sky is
demonstrated by the fact that 4212 (50\%) of the galaxies belong to
the SSC and its inmediate neighborhood, if we consider as such all
galaxies with velocities in the range $9000-18000~\kms$ (a total
depth of $90~h^{-1}$~Mpc). It can be seen that by probing large
regions of the SSC away from the richer Abell clusters, we have
confirmed significant additional structures which make complex links
with the main cluster locations. The transverse dimensions of the
area surveyed are defined from R.A. $\simeq 12^{\rm h}40^{\rm m}$ to
$\simeq 14^{\rm h}10^{\rm m}$ and from Dec $\simeq -24^\circ$ to
$\simeq -38^\circ$, corresponding to at least $40 \times
25~h^{-1}~{\rm Mpc}$.

\begin{figure*}[htb]
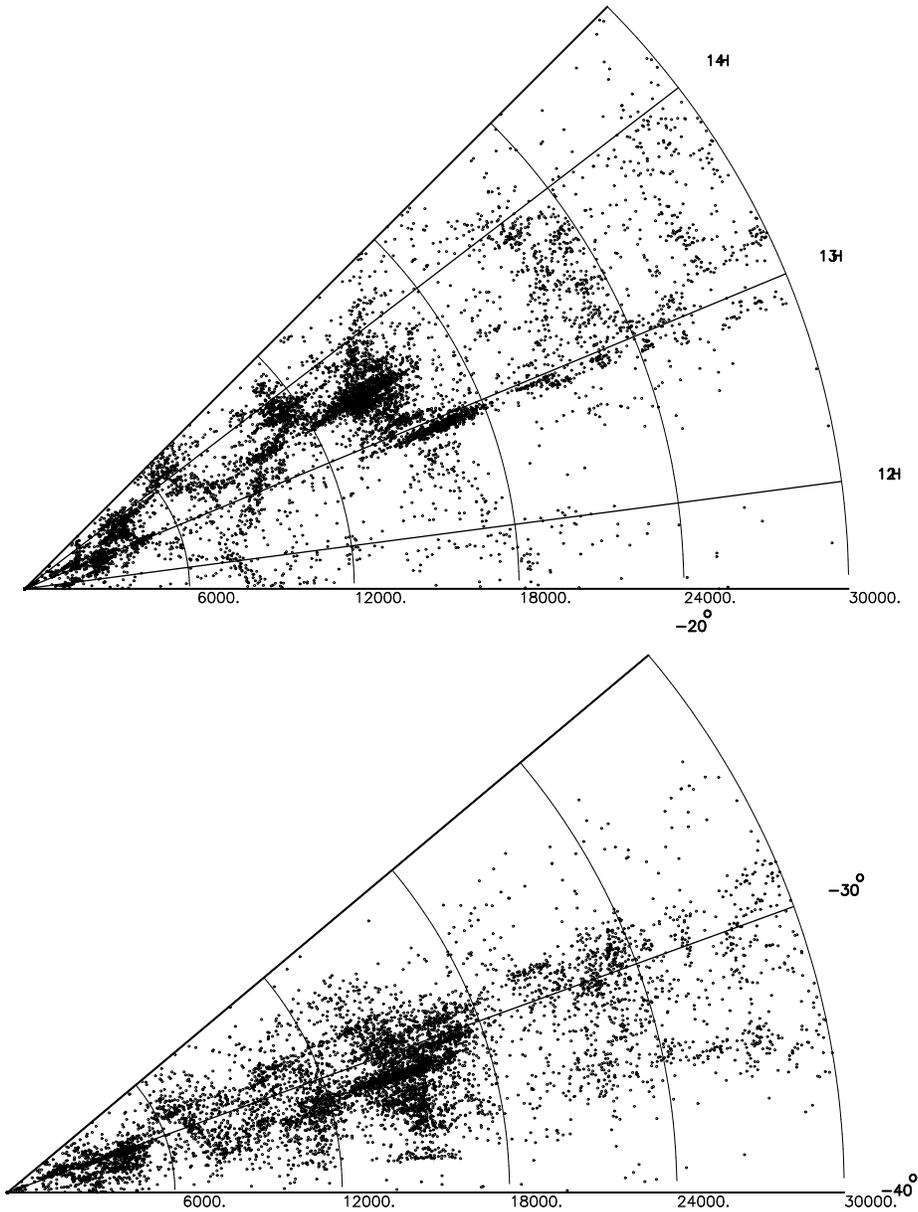

\centering
\includegraphics[width=8.0cm,angle=-90]{smallconshap.ps}
\includegraphics[width=8.0cm,angle=-90]{smalldelshap.ps}
\caption{Two projections of the distribution of galaxies with
measured redshifts in the region of the Shapley Supercluster. The
radial coordinate, the recession velocity determined from the
redshift, $cz$, measured in $\kms$, is an imperfect surrogate (see
text) for the galaxy's distance to us, who would be at the vertex.
The angle in declination is expanded by a factor~2 relative to its
true size for clarity.}
\end{figure*}

\subsection{Structure of the Shapley Supercluster}

The larger velocity catalogue used in this analysis confirms the
general structure and main features of the SSC already discussed in
Quintana \etal (2000). For completeness, we summarize them here,
pointing out new features that have surfaced. The central region
(CR) is roughly spherical in shape and has at its core the
highest-density, elongated volume containing the Abell clusters
A3562, A3558 and A3556, with almost identical recession velocities
around $14400\kms$, and the groups SC1329-314 and SC1327-312, whose
more discrepant velocities (by several hundred kilometers per
second) could be attributed to the infall component along the line
of sight. Towards the south of the elongated feature, the central
region contains also the cluster A3560. As described in Reisenegger
\etal (2000), the whole of this central region and all of its
immediate surroundings are within the volume that is currently
undergoing gravitational collapse.

We note the presence of a prominent foreground wall of galaxies
(Hydra-Centaurus region) at $\overline{v}= 4000\kms$. This
distribution can be related to the nearby cluster A3627, associated
with the ``Great Attractor'' (Kraan-Korteweg \etal 1996). Moreover,
a remarkable bridge of galaxies, groups and clusters, the so-called
``Front Eastern Wall" (Quintana \etal 2000), extends to the east and
in front of the supercluster, the densest part being at $\simeq
10000-11000\kms$, located to the east. It contains the clusters
A3571, A3572, A3575 and the group SC1336-314. The A3570 cluster is
located at the southern tip of the observed part of the wall and
A3578 at its northern one. This wall establishes a link between the
Hydra-Centaurus region and the SSC, while a second one extends
towards the west at $\overline{v}=8300\kms$. Clumps of objects
clearly link the two main structures. However, care must be taken in
the interpretation of the wedge plots because of the finger-of-God
effect evident in the main SSC concentrations (made especially
prominent by the higher completeness in these cluster regions) and
because of an analogous effect (with opposite sign) due to the
inflow on larger scales, which makes the overdensities appear more
overdense in redshift space than they are in the real space.
Furthermore, the opening angle of the wedge diagram on the lower
panel of Fig. 3 is enhanced by a factor of two, so the structures
are stretched across the line of sight.

As is well known, a large concentration of galaxies and clusters at
about R.A.$= 12^{\rm h}50^{\rm m}$ and around $v=16000-17000\kms$,
the ``A3528 complex'' (Quintana \etal 1995, Bardelli \etal 2000) is
connected to the main body of the SSC by a broad bridge of galaxies.
It can also be seen from the wedge diagram in declination (Fig.~3,
lower panel) that the Southern part of the SSC consists of two large
sheets of galaxies where the known Abell clusters represent the
peaks of maximum density. In this diagram, the more distant sheet,
in particular, at $\overline{v}=15000\kms$ is present right across
the observed region from $-26^{\circ}$ up to $-38^{\circ}$, so the
true extent of this wall is still currently unknown. The Southern
part of this wall may be an extension of the plane of galaxies
defined by Bardelli \etal (2000), since it has the same offset of
$-5h^{-1}$~Mpc when analysed by Bardelli \etal in their Figure~4.

\subsection{Connections of the SSC to surrounding superclusters}

Figure 3 also shows the presence of a large background complex of
galaxies at $\overline{v}= 22500\kms$, as well as more distant
clusters. This complex is associated with the supercluster SCL134 in
the catalogue of Einasto et al. (2001). Background structures appear
to be linked to the SSC by an extremely long and thin filament
starting at $R.A.=13^{\rm h}$. In Fig.~4, this filament appears to
extend out to $v= 48000\kms$, although its reality cannot be fully
ascertained, given the non-uniform selection of galaxies and
possible projection effects. However, two parts of this filament
correspond to Einasto's superclusters SCL127 at
$\overline{v}=22700\kms$ and SCL129 at $\overline{v}=28300 \kms$.
The two superclusters SCL146 and SCL266 also seem to be associated
with radial extensions of the SSC.

The Shapley Supercluster is clearly linked to other huge
superstructures, as shown in the Figures 9 and 10 of Jones et al.
(2004). Apart from the radial connection to the Hydra-Centaurus
complex, a tangent bridge of galaxies extends in the direction of
the Sextans supercluster at $z\simeq 0.04$. This distribution of
galaxies across the southern sky, projected across the full range of
southern declinations and each of the individual $10^\circ$ ranges
in Jones et al. (2004), shows the very large number of ramifications
connecting already identified structures between R.A.$= 10^{\rm h}$
and $15^{\rm h}$.

\begin{figure*}[htb]
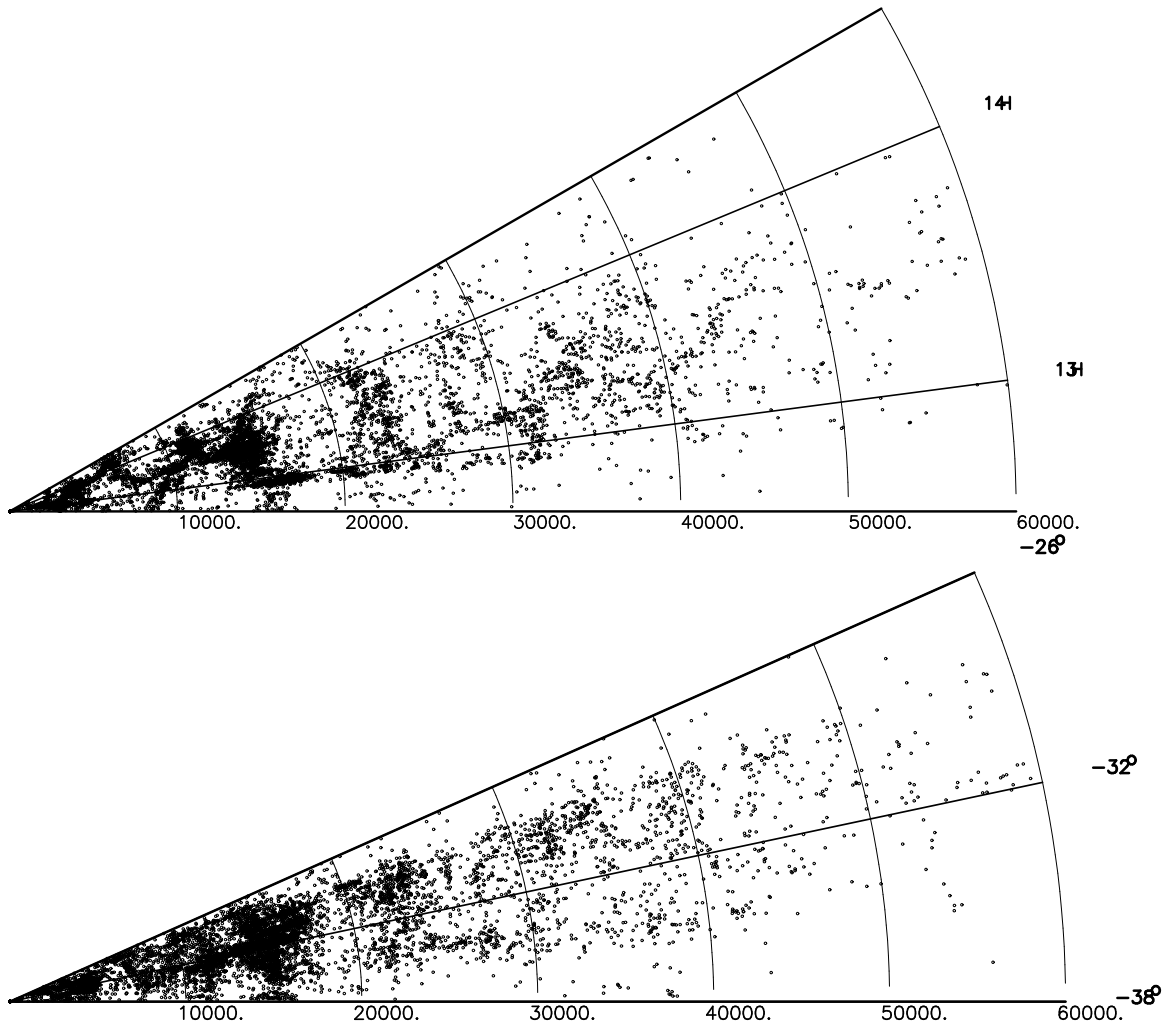

\centering
\includegraphics[width=7.0cm,angle=-90]{conshap.ps}
\includegraphics[width=6.5cm,angle=-90]{delshap.ps}
\caption{Wedge diagrams in right ascension and declination extending
out to $v= 60000\kms$, which suggest the presence of a long, thin
filament starting at the SSC at ${\rm R.A.}= 13^{\rm h}$ and
extending out to $v=48000\kms$.}
\end{figure*}

\section{Clusters of galaxies in the direction of the Shapley Supercluster}

Velocity dispersions are an essential piece of information to study
cluster dynamics, because they directly probe the cluster potential.
For a preliminary study (to be refined in Quintana et al. 2005), we
searched the NED/IPAC Extragalactic Database for all clusters in the
area defined by $12^{\rm h}30^{\rm m} <$ R.A. $< 14^{\rm h}30^{\rm
m}$ and $-20^\circ >$ Dec $>-40^\circ$ (30 degrees by 20 degrees on
the sky). For clusters with known velocities, we limited our search
to $cz<40000\kms$. The output list contains 141 clusters, 134 of
which are listed in the ACO catalog, four are previously identified
groups of galaxies (Quintana et al. 2000), two are X-ray detected
clusters (RXC J1304.2-3030 by B\"{o}hringer at al. 2004, RX
J1252.5-3116 by Pierre et al 1994), and one, CL 1322-30, is a
cluster detected by Stein (1996).

The cluster centers were chosen by one of the following criteria:
(a) from published X-ray centers; (b) from the position of the
brightest cluster galaxy (cD or D if present); (c) using the Abell
cluster center. For clusters with known velocities, most of the
centers listed in the NED correspond to the bulk of the Shapley
Supercluster and are given by Quintana et al. (2000). For all
clusters with unknown velocities, we used the Abell cluster center.

As a relatively rough, first approach, we examined the velocity
distribution of galaxies within a common angular radius of 0.5
degree of each cluster center (corresponding to $1.3~h^{-1}{\rm
Mpc}$, roughly an Abell radius, at $z=0.05$, but smaller or larger
for foreground or background clusters, respectively). The average
recession velocity, $\langle V \rangle$, and the one-dimensional
line-of-sight velocity dispersion, $\sigma_V$, were calculated using
the bi-weighted estimators of location and scale of Beers et al.
(1990). We used an iterative procedure by calculating the location
and scale using the ROSTAT program and applying a $3\sigma$-clipping
algorithm to the results. We repeated this procedure until the
velocity dispersion converged to a constant value (within two or
three iterations).

The number of clusters with at least six galaxies in our velocity
sample is 68. Forty-four clusters are in the velocity range of the
Shapley Supercluster ($9000 \kms < cz < 18000 \kms$) and eight are
foreground. In the velocty interval of $18000 \kms < cz < 30000$, we
identified 11 clusters. The values of $\langle V \rangle$ and
$\sigma_{V}$ for individual clusters are presented in Table~4.
$N_{sel}$ is the number of galaxies selected within 0.5 degree and
$N_{mem}$ is the number of member galaxies. The identification of
the cluster as a member of a particular supercluster is indicated in
the comments column with the designation of Einasto \etal (2001).
SCL124 and SCL128 correspond, respectively, to the SSC and the
Hydra-Centaurus supercluster.

Many clusters present significant substructures, identified by the
multi-modality of their velocity distributions. In some cases, the
main structures associated with catalogued clusters are behind (e.g.
A3524, A3531, A3535, AS717, A3546, AS725, A3549, A3551, A3557,
SC1340-294, A3576) or in front of (e.g. RXC J1304.2-3030, CL1322-30,
A3565, A3574, AS753, A3581, AS761) the Shapley supercluster.

Most of the clusters are well isolated in their three-dimensional
distribution, with some exceptions. Four clusters in the sample
present significant foreground and background structures (A3542,
A3545, AS724, AS731). For these clusters, $\langle V \rangle$ and
$\sigma_{V}$ are reliable values, however a more detail analysis is
required to determine: (a) where the cluster is located; (b) if the
foreground and background structures seen in the 3-D distribution
are real.

The distribution on the sky of the clusters and galaxies in the
velocity range of the Shapley Supercluster is presented in Fig.~5.
At  $\simeq 9000-13000\kms$, two main cluster clumps are present
(left panel). The first one is formed by the clusters A3571, A3572,
A3575m AS748 and AS744 (the ``Front Eastern Wall'', see section 4.2)
and the second is formed by the clusters A1736a and AS736. At
$\simeq 13000-18000\kms$, it is possible to see three main cluster
clumps: the core of the SSC, formed by the clusters A3552, A3556,
A3558, and A3562 and the groups SC1329-314 and SC1327-312, and the
two extensions formed by A3568, A3566, and A3464 and by A3528,
A3530, and A3532, respectively (see section 4.2).

\begin{figure*}[htb]
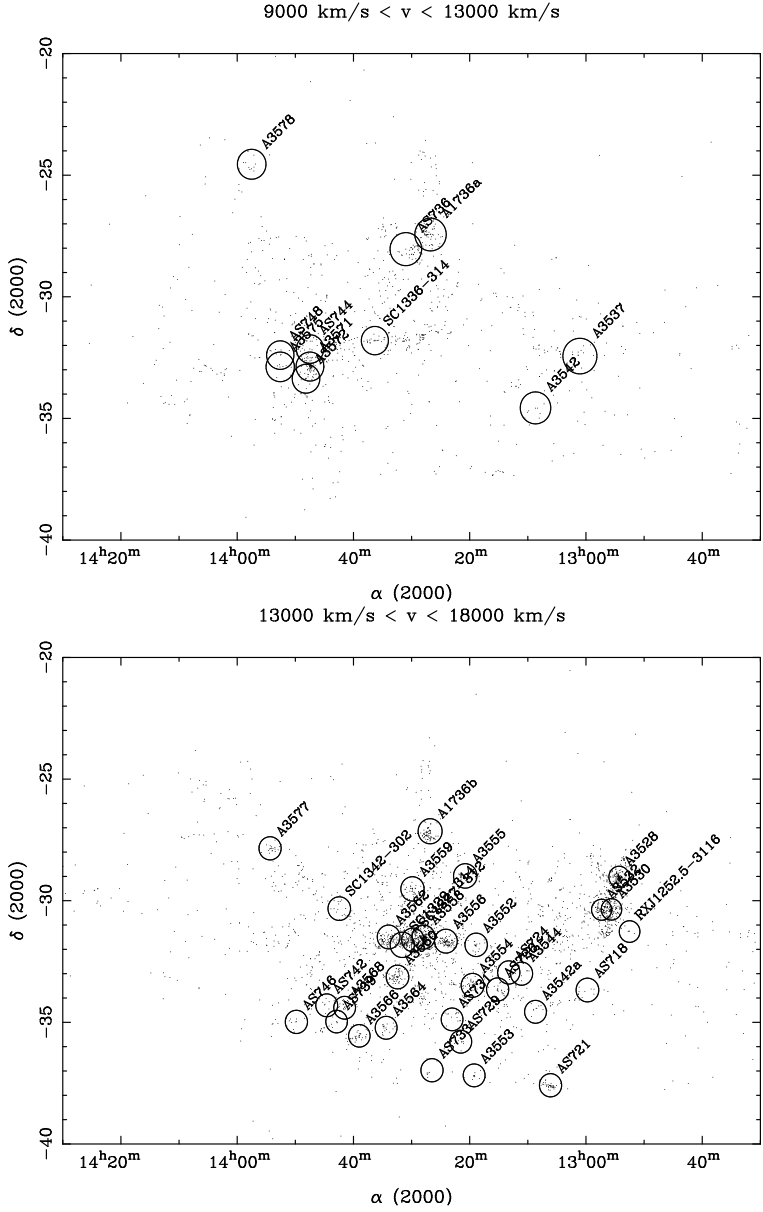

\centering
\includegraphics[width=8.0cm,angle=-90]{ssc9000_13000.ps}
\vspace{1cm}
\includegraphics[width=8.0cm,angle=-90]{ssc13000_18000.ps}

\caption{Distribution on the sky of the 44 clusters and galaxies in
the velocity range of the SSC. Upper panel: cluster with velocities
between $9000\kms$ and $13000\kms$. Lower panel: clusters between
$13000\kms$ and $18000\kms$}
\end{figure*}

\section{Global properties}

Ignoring redshift-space distortions due to peculiar velocities, the
surveyed area of $225\deg^2$, from 9000 to $18000\kms$, corresponds
to a spatial volume $V_{\rm SSC}=1.17\times 10^5h^{-3}{\rm Mpc}^3$,
equivalent to that of a sphere of radius $r_{\rm eff}=30.3h^{-1}{\rm
Mpc}$. This is near the middle of the range of supercluster sizes
found by Einasto et al. (2003a,b) in their systematic searches of
the Sloan Digital Sky Survey (SDSS) and Las Campanas Redshift Survey
(LCRS) data. However, we note that their density threshold for
selecting superclusters is 1.8 times the average redshift-space
density, much lower than the overdensity of 5.4 we find in our
study. This means that the true extension of the Shapley
Supercluster, applying the criterion of Einasto and collaborators,
would be much larger than the area of our study. In the extreme case
of having an empty volume around the region of our study, the radius
enclosing an average overdensity of 1.8 is $(5.4/1.8)^{1/3}r_{\rm
eff}\simeq 44~h^{-1}{\rm Mpc}$, increasing to $\simeq 54~h^{-1}{\rm
Mpc}$ if the surrounding volume is assumed to be of average density,
which is probably still an underestimate.

An estimate of the total luminosity, based on the assumption that
the surveyed galaxies are distributed as the total light, can be
given as
\begin{equation}
L_{\rm tot}=({\rm overdensity})\times{\bar j}\times V_{\rm
SSC}\simeq 1.4\times 10^{14}h^{-2}L_\odot,
\end{equation}
where the average $b_j-$band luminosity of the Universe, ${\bar
j}\simeq 2.3\times 10^8hL_\odot{\rm Mpc}^{-3},$ was obtained from
the 2dF luminosity function of Norberg et al. (2002), and
corresponds to an average mass-to-light ratio of $\simeq
360(\Omega_{\rm M}/0.3)h(M_\odot/L_\odot)$. An alternative estimate,
strictly speaking a lower bound, is the ``observed luminosity'',
namely the sum of the luminosities of all the galaxies in our
catalogue within the specified volume,
\begin{equation}
L_{\rm obs}=2.2\times 10^{13}h^{-2}L_\odot,
\end{equation}
about 7~times smaller than our estimate for $L_{\rm tot}$. This is
not surprising, as our catalogue is not complete, and the total
luminosity is strongly dominated by a few of the brightest galaxies,
some of which may have been excluded from the surveys under the
assumption that they were foreground objects, not belonging to the
SSC.

These values, although still underestimates for the true luminosity
of the SSC, because of the limited well-sampled volume, are higher
than any of the corresponding luminosities of the superclusters in
the SDSS-South and comparable to the brightest in the SDSS-North and
the LCRS, which are at substantially higher redshifts than the SSC
(Einasto et al. 2003a,b). This confirms that the SSC is a truly
exceptional structure, as found by previous surveys of the local
Universe (e.g., Einasto et al. 2001).

Similarly, we can assume that the surveyed galaxies trace mass and
estimate the total mass within the surveyed volume as
\begin{equation}
M_{\rm tot}=({\rm overdensity})\times \Omega_{\rm M}\rho_{\rm
crit}V_{\rm SSC}=5\times 10^{16}h^{-1}M_\odot,
\end{equation}
where $\rho_{\rm crit}$ is the critical density for a flat Universe,
and we assumed that the standard mass-density parameter is
$\Omega_{\rm M}=0.3$. A lower bound on this mass, in essentially the
same volume, was obtained by Ragone et al. (2005), who identified
galaxy systems and estimated their individual masses. Adding these,
they find $M_{\rm obs}=4.8\times 10^{15}h^{-1}M_\odot$. Correcting
(based on simulations) for systems presumably missed in the survey
(mainly because of non-uniform incompleteness), this value increases
to $M_{\rm corr}=1.1\times 10^{16}h^{-1}M_\odot$, about 1/4 of our
$M_{\rm tot}$. If real and not due to biases of one or the other
method, this would indicate that in fact a large fraction of the SSC
(80\%, if taken at face value) is not in galaxy clusters or groups.

Another dynamical estimate for the mass was given by Reisenegger
\etal (2000), who searched for ``caustics'' in the projected
radius-redshift diagram of the SSC and applied to these both a
straightforward spherical collapse model (Reg\"os \& Geller 1989)
and the heuristic escape-velocity method of Diaferio (1999). The
mass within a radius $r=8 h^{-1}$ Mpc was found to be $\sim 10^{16}
h^{-1} M_\odot$ in the former, and $\sim 2\times 10^{15}h^{-1}
M_\odot$ in the latter, corresponding to overdensities
$\rho/(\Omega_{\rm M}\rho_{\rm crit})\simeq 60$ and 10,
respectively. A lower limit on the virialized mass in clusters
turned out to be similar to the latter result. These results, based
on a much smaller volume, are therefore hard to compare, but do not
seem inconsistent with those of the present study.

The mass required at the distance of the SSC to produce the observed
motion of the Local Group with respect to the CMB is $M_{\rm
dipole}\approx (1-3) \times 10^{17}(\Omega_{\rm
M}/0.3)^{0.4}h^{-1}M_\odot$ (Hoffman \etal 2001), not very far above
the total mass of the SSC as estimated here. In consistent models
(e.g. Branchini \etal 1999) of the density and velocity distribution
on large scales (where density fluctuations are small) in the local
Universe, the SSC figures prominently, although the Local Group
motion originates from a combination of several attractors.

\section{Conclusions}
This paper is based on the largest velocity catalogue available up
to now to analyse various properties of the Shapley Supercluster. It
shows a completeness analysis for 8632 galaxies with measured
velocities (10529 velocity measurements), including 1201 new
velocities for 1166 galaxies to be published in Quintana \etal
(2005). The completeness is highest for the brighter magnitude
limits, peaking at $78\%$ for the whole sample at $b_{j}<17.0$. The
galaxy overdensity in the full SSC region is $\sim 5.4$, and in the
restricted intercluster region it is $\sim 3.8$. These results place
the SSC among the largest of superclusters found in recent,
systematic surveys at higher redshifts. Its inferred mass is large
enough to have a non-negligible effect on the observed Local Group
motion. The SSC has a general flat morphology, extending further
than all previous estimates and linking the foreground
Hydra-Centaurus region. We also presented and discussed a
preliminary catalogue of 68 galaxy clusters with dynamical
information in this area of the sky, 44 of which are in the redshift
range of the SSC.

\acknowledgements{This research was partially supported by the
cooperative programme ECOS/CONICYT C96U04. DP and ES thank the
Astronomy Department at PUC (Santiago de Chile) for its hospitality.
DP acknowledges receipt of a France-Australie PICS in support of
visits to Siding-Spring Observatory and the PICS-CNPq France-Brazil
cooperation 910068/00-3 in support of visits to the University of
Sao Paulo (IAG/USP). HQ was supported by the FONDAP Centre for
Astrophysics. ERC is supported by the Gemini Observatory, which is
operated by the Association of Universities for Research in
Astronomy Inc., on behalf of the international Gemini partnership of
Argentina, Australia, Brazil, Canada, Chile, the United Kingdom and
the United States of America. AR and RD received support from
FONDECYT through its Regular Project 1020840. We thank our referee,
Prof. J. Einasto, for his precious remarks and suggestions. Some of
the results in this paper have been derived using the HEALPix
package (G\'orski \etal 2005).}

\newpage

\begin{longtable}{lccrrlll}
\caption{\label{tab.tab4} Clusters of galaxies in the area of the
Shapley Supercluster}\\
\hline\hline Cluster & R.A. (J2000) & Dec (2000) & $N_{\rm sel}$ &
$N_{\rm mem}$ &
$\langle V \rangle$ & $\sigma_{V}$ & Comments \\
  & $^{h}$ $^{m}$ $^{s}$ & $^{\circ}$ $^{'}$ $^{''}$ &  &  &
$\kms$ & $\kms$ &  \\
(1) & (2) & (3) & (4) & (5) & (6) & (7) & (8) \\
\hline
\endfirsthead
\caption{continued}\\
\hline\hline Cluster & R.A. (J2000) & Dec (J2000) & $N_{\rm sel}$ &
$N_{\rm mem}$ & $\langle V \rangle$ & $\sigma_{V}$ & Comments \\
  & $^{h}$ $^{m}$ $^{s}$ & $^{\circ}$ $^{'}$ $^{''}$ &  &  &
$\kms$ & $\kms$ &  \\
(1) & (2) & (3) & (4) & (5) & (6) & (7) & (8) \\
\hline
\endhead
\hline
\endfoot
A3524     & 12 40 05.3 & --34 13 28 &  19 &   9 & 22009$\pm$171 & 465$\pm$165 & background \\
          &            &            &     &     &               &         & \\
RX J1252.5-3116 & 12 52 30.0 & --31 16 00 &  11 &  6 & 16110$\pm$206 & 425$\pm$102 & X-ray center (Pierre at al. 1994) \\
          &            &            &     &     &               &         & \\
A3528     & 12 54 18.2 & --29 01 21 & 198 &  53 & 16294$\pm$100 & 729$\pm$109 & SCL124 \\
          &            &            &     &     &               &         & \\
A3530     & 12 55 36.9 & --30 21 20 &  66 &  35 & 16253$\pm$100 & 583$\pm$65 & $r_{\rm sel}<13^{'}$ SCL124 \\
          &            &            &     &     &               &         & \\
A3531     & 12 57 08.2 & --32 55 19 &  33 &  17 & 22779$\pm$102 & 404$\pm$115 & SCL127 \\
          &            &            &     &     &               &         & \\
A3532     & 12 57 19.2 & --30 22 18 &  67 &  60 & 16667$\pm$87  & 670$\pm$86  & $r_{\rm sel}<13^{'}$ SCL124 \\
          &            &            &     &     &               &         & \\
A3535     & 12 57 48.6 & --28 29 17 & 105 &  56 & 20213$\pm$58  & 436$\pm$26 & background \\
          &            &            &     &     &               &         & \\
AS717     & 12 58 04.7 & --28 11 11 &  97 &  10 & 22668$\pm$124 & 362$\pm$89 & background \\
          &            &            &     &     &               &         & \\
AS718     & 12 59 45.0 & --33 40 15 &  73 &  14 & 14527$\pm$68  & 243$\pm$50  & \\
          &            &            &     &     &               &         & \\
A3537     & 13 01 02.7 & --32 26 14 &  44 &  12 &  9371$\pm$76  & 248$\pm$133  & SCL128 \\
          &            &            &     &     &               &         & \\
RXC J1304.2-3030 & 13 04 16.7 & -30 30 55 &  52 &  29 & 3318$\pm$64  & 331$\pm$36  & X-ray (B\"{o}hringer \etal 2004) \\
          &            &            &     &     &               &         & foreground \\
AS721     & 13 06 06.0 & --37 35 11 &  43 &  36 & 14876$\pm$116 & 686$\pm$72  & \\
          &            &            &     &     &               &         & \\
A3542     & 13 08 41.0 & --34 33 59 &  46 &   6 & 10528$\pm$49  & 373$\pm$84  & SCL124 \\
A3542a    &            &            &     &   7 & 15146$\pm$100 & 233$\pm$44  & \\
A3542b    &            &            &     &  13 & 27087$\pm$252 & 859$\pm$150 & background \\
          &            &            &     &     &               &         & \\
A3544     & 13 11 04.7 & --32 59 56 &  26 &  11 & 14969$\pm$164 & 502$\pm$140  & \\
          &            &            &     &     &               &         & \\
A3545     & 13 11 23.3 & --34 04 57 &  49 &  13 & 29879$\pm$142 & 484$\pm$112 & SCL129C \\
A3545a    &            &            &     &   8 & 27190$\pm$165 & 411$\pm$137 & \\
          &            &            &     &     &               &         & \\
A3546     & 13 13 03.7 & --29 58 55 &  44 &  14 & 32113$\pm$77  & 275$\pm$40  & SCL129C \\
          &            &            &     &     &               &         &\\
AS724     & 13 13 17.0 & --32 56 55 &  39 &  12 & 14864$\pm$157 & 510$\pm$85  & \\
AS724a    &            &            &     &  17 & 30401$\pm$150 & 597$\pm$105 & background \\
          &            &            &     &     &               &         &\\
AS725     & 13 14 09.8 & --30 11 53 &  18 &   9 & 32172$\pm$92  & 250$\pm$80  & background \\
          &            &            &     &     &               &         & \\
A3549     & 13 14 21.5 & --29 26 53 &  36 &   8 & 22878$\pm$173 & 432$\pm$133 & SCL127 \\
A3549b    &            &            &     &  11 & 31300$\pm$174 & 953$\pm$114 & \\
          &            &            &     &     &               &         & \\
AS726     & 13 15 11.7 & --33 38 52 &  34 &  19 & 14892$\pm$137  & 578$\pm$77  & \\
          &            &            &     &     &               &         & \\
A3551     & 13 18 10.8 & --30 55 46 &  21 &  12 & 37434$\pm$294 & 952$\pm$243 & background \\
          &            &            &     &     &               &         & \\
A3552     & 13 18 53.5 & --31 48 48 &  47 &  34 & 14753$\pm$119 & 682$\pm$60  & SCL124 \\
          &            &            &     &     &               &         & \\
A3553     & 13 19 14.6 & --37 10 45 &  20 &  12 & 15141$\pm$153 & 497$\pm$95  & SCL124 \\
          &            &            &     &     &               &         & \\
A3554     & 13 19 30.5 & --33 28 45 &  61 &  36 & 14431$\pm$94  & 560$\pm$66  & SCL124 \\
          &            &            &     &     &               &         & \\
A3555     & 13 20 46.2 & --28 58 47 &  45 &  22 & 14034$\pm$47  & 214$\pm$33  & SCL124 \\
          &            &            &     &     &               &         & \\
AS729     & 13 21 32.2 & --35 47 44 &  55 &  20 & 15158$\pm$106 & 462$\pm$102 & \\
          &            &            &     &     &               &         & \\
AS731     & 13 23 01.9 & --34 52 41 & 104 &  38 & 15230$\pm$84  & 514$\pm$61  & \\
AS731a    &            &            &     &  17 &  7940$\pm$85  & 339$\pm$69  & \\
          &            &            &     &     &               &         & \\
A3556     & 13 24 06.2 & --31 39 44 & 123 &  75 & 14439$\pm$71  & 618$\pm$50  & SCL124 \\
          &            &            &     &     &               &         & \\
CL1322-30 & 13 24 47.6 & --30 17 38 &  56 &  21 &  4242$\pm$64  & 287$\pm$46  & detected by Stein (1996)\\
          &            &            &     &     &               &         &     foreground \\
A3557     & 13 24 52.8 & --28 52 43 &  59 &  22 & 23401$\pm$75  & 343$\pm$73  &  SCL 127 \\
          &            &            &     &     &               &         & \\
AS733     & 13 26 29.3 & --36 58 07 &  22 &   8 & 14908$\pm$167 & 417$\pm$77  & \\
          &            &            &     &     &               &         & \\
A1736a    & 13 26 44.3 & --27 26 22 & 264 &  92 & 10215$\pm$47  & 450$\pm$30  & \\
          &            &            &     &     &               &         & \\
A1736b    & 13 26 48.7 & --27 08 38 & 264 & 143 & 13654$\pm$73  & 873$\pm$52  & \\
          &            &            &     &     &               &         & \\
A3558     & 13 27 56.9 & --31 29 44 & 285 & 247 & 14307$\pm$64  & 1010$\pm$44 & SCL124 \\
          &            &            &     &     &               &         & \\
SC1327-312& 13 29 47.0 & --31 36 29 &  48 &  35 & 14650$\pm$147 & 857$\pm$117 & X-ray center (Breen et al.1994) \\
          &            &            &     &     &               &         & $r_{\rm sel}<11^{'}$ \\
          &            &            &     &     &               &         & \\
A3559     & 13 29 51.0 & --29 30 51 & 168 &  82 & 14130$\pm$57  & 519$\pm$45  & SCL124 \\
          &            &            &     &     &               &         & \\
AS736     & 13 30 59.2 & --28 02 26 &  64 &  30 & 10140$\pm$95  &  514$\pm$70 & \\
          &            &            &     &     &               &         & \\
SC1329-314& 13 31 36.0 & --31 48 46 &  38 &  35 & 13541$\pm$151 & 883$\pm$168 & X-ray center (Breen et al.1994)\\
          &            &            &     &     &               &         & $r_{\rm sel}<9^{'}$ \\
          &            &            &     &     &               &         & \\
A3560     & 13 32 25.3 & --33 08 12 &  73 &  56 & 14551$\pm$106 & 793$\pm$116 & SCL124 \\
          &            &            &     &     &               &         & \\
A3562     & 13 33 56.8 & --31 29 23 &  41 &  40 & 14455$\pm$191 & 1197$\pm$194& $r_{\rm sel}<12^{'}$ SCL124 \\
          &            &            &     &     &               &         & \\
A3564     & 13 34 22.4 & --35 13 22 &  64 &  30 & 15116$\pm$73  & 393$\pm$64  & SCL124 \\
          &            &            &     &     &               &         & \\
SC1336-314& 13 36 19.0 & --31 48 00 &  85 &   7 & 11982$\pm$189 & 941$\pm$261 & \\
          &            &            &     &     &               &         & \\
A3565     & 13 36 39.8 & --33 58 18 &  27 &  13 &  3860$\pm$43  & 148$\pm$118 & SCL128 \\
      &            &            &     &     &               &         & \\
A3566     & 13 38 59.4 & --35 33 13 &  42 &  25 & 15388$\pm$106 & 519$\pm$58  & SCL124 \\
          &            &            &     &     &               &         & \\
SC1340-294& 13 40 00.0 & --29 45 59 &  26 &  15 & 23226$\pm$126 & 467$\pm$75  & background \\
          &            &            &     &     &               &         & \\
A3568     & 13 41 31.0 & --34 24 00 &  39 &  27 & 15218$\pm$143 & 726$\pm$101 & \\
          &            &            &     &     &               &         & \\
SC1342-302& 13 42 26.4 & --30 19 26 &  38 &  14 & 14621$\pm$105 & 375$\pm$49  & New cluster \\
          &            &            &     &     &               &         & \\
AS739     & 13 42 53.7 & --34 58 06 &  29 &   8 & 15318$\pm$243 & 607$\pm$134 & \\
          &            &            &     &     &               &         & \\
AS742     & 13 44 36.0 & --34 18 00 &  53 &  22 & 15124$\pm$100 & 455$\pm$93  & \\
          &            &            &     &     &               &         & \\
AS744     & 13 47 28.4 & --32 08 57 &  48 &  31 & 11882$\pm$138 & 756$\pm$118 & \\
          &            &            &     &     &               &         & \\
A3571     & 13 47 28.9 & --32 51 58 &  95 &  92 & 11676$\pm$106 &1016$\pm$80  & SCL124 \\
          &            &            &     &     &               &         & \\
A3572     & 13 48 09.7 & --33 22 05 &  14 &   7 & 11740$\pm$279 & 651$\pm$278 & $r_{\rm sel}<13^{'}$ SCL124 \\
          &            &            &     &     &               &         & \\
A3574     & 13 49 09.4 & --30 17 55 &  57 &  38 &  4512$\pm$90  & 549$\pm$56  & SCL128 \\
          &            &            &     &     &               &         & \\
AS746     & 13 49 49.0 & --34 58 52 &  35 &  12 & 14877$\pm$123 & 399$\pm$92  & \\
          &            &            &     &     &               &         & \\
AS748     & 13 52 35.5 & --32 23 48 &  27 &  13 & 11866$\pm$97  & 332$\pm$97  & \\
          &            &            &     &     &               &         & \\
A3575     & 13 52 35.8 & --32 52 47 &  39 &  20 & 11405$\pm$122 & 531$\pm$92  & SCL124 \\
          &            &            &     &     &               &         & \\
A3576     & 13 52 45.9 & --30 17 47 &  51 &  21 & 22114$\pm$64  & 283$\pm$42  & background \\
          &            &            &     &     &               &         & \\
A3577     & 13 54 20.4 & --27 50 45 &  37 &  24 & 14890$\pm$131 & 628$\pm$98  & SCL124 \\
          &            &            &     &     &               &         & \\
A3578     & 13 57 30.2 & --24 33 05 &  16 &   9 & 11297$\pm$150 & 407$\pm$79  & SCL124 \\
          &            &            &     &     &               &         & \\
          &            &            &     &     &               &         &   \\
AS753     & 14 03 38.5 & --33 58 24 &  19 &  15 &  3960$\pm$84  & 312$\pm$143 & foreground \\
          &            &            &     &     &               &         & \\
A3581     & 14 07 27.5 & --27 01 16 &  32 &  24 &  6725$\pm$120 & 574$\pm$56  & SCL128 \\
          &            &            &     &     &               &         & \\
AS761     & 14 18 47.2 & --27 25 47 &  17 &  12 &  6902$\pm$153 & 495$\pm$115 & foreground \\
          &            &            &     &     &               &         & \\
\end{longtable}


\begin{thebibliography}{}

\bibitem[]{Baldi01} Baldi A., Bardelli, S., Zucca E. 2001, MNRAS, 324 509.

\bibitem[]{Bardelli94} Bardelli S., Zucca E., Vettolani G., Zamorani G.,
Scaramella R. \etal 1994, MNRAS, 267, 665.

\bibitem[]{Bardelli96} Bardelli S., Zucca E., Malizia A., Zamorani G.
\etal 1996, A\&A, 305, 435.

\bibitem[]{Bardelli98} Bardelli S., Zucca E., Zamorani G., Vettolani G.,
Scaramella R. 1998, MNRAS, 296, 599.

\bibitem[]{Bardelli00} Bardelli S., Zucca E., Zamorani G., Moscardini L.,
Scaramella, R. 2000, MNRAS, 312, 540

\bibitem[]{Bardelli01} Bardelli S., Zucca E., Baldi A., 2001 MNRAS, 320, 387.

\bibitem[]{Branchini99} Branchini, E., Teodoro, L., Frenk, C. S., Schmoldt, I.,
Efstathiou, G. \etal 1999, MNRAS, 308, 1.

\bibitem[]{Bohringer04} B\"{o}hringer, H., Schuecker, P., Guzzo, L.,
Collins, C. A., Voges, W., Cruddace, R. G. \etal 2004, A\&A, 425, 367

\bibitem[]{Breen01} Breen J., Raychaudhury S., Forman W., Jones C.
1994, ApJ, 424, 59.

\bibitem[]{Colless01} Colless M.M. \etal (2dFGRS team) 2001, MNRAS, 328, 1039.

\bibitem[]{Diaferio99} Diaferio, A. 1999, MNRAS, 309, 610.

\bibitem[]{Drink98} Drinkwater M.J., Proust D., Parker Q.A., Quintana H.,
Slezak E. 1998, 14th IAP conference ``Wide Field Surveys in
Cosmology'', Paris, 26-29 May 1998, ed. S. Colombi, Y. Mellier \& B.
Raban, Editions Fronti\`eres, p. 392.

\bibitem[]{Drink99} Drinkwater M.J., Proust D., Parker Q.A., Quintana H.,
Slezak E. 1999, PASA, 16, 113.

\bibitem[]{Drink04} Drinkwater M.J., Parker Q.A., Proust D., Slezak E.,
Quintana H. 2004, PASA, 21, 89.

\bibitem[]{Einasto01} Einasto M., Einasto J., Tago E., Muller V., Andernach H.
2001, AJ 122, 2222.

\bibitem[]{Einasto03a} Einasto J., H\"utsi G., Einasto M.,
Saar E., Tucker D.L. \etal 2003a, A\&A, 405, 425.

\bibitem[]{Einasto03b} Einasto J., Einasto M., H\"utsi G.,
Saar E., Tucker D.L. \etal 2003b, A\&A, 410, 425.

\bibitem[]{Einasto05} Einasto J. 2005, private communication.

\bibitem[]{Felenbok97} Felenbok, P., Guerin, J., Fernandez, A., Cayatte, V.,
Balkowski \etal 1997, Experimental Astronomy, 7, 65.

\bibitem[]{Fleenor05} Fleenor M.C., Rose J.A., Christiansen W.A.,
Hunstead R.W. Johnson-Hollit M. \etal 2005, astro-ph/0505361.

\bibitem[]{Gorski05} G\'orski, K. M., Hivon, E., Banday, A. J., Wandelt, B. D.,
Hansen \etal 2005, ApJ, 622, 759.

\bibitem[]{Hambly01a} Hambly N.C., Mac Gillivaray H.T., Read M.A.,
Tritton S.B., Thomson E.B. \etal 2001a, MNRAS, 326, 1279.

\bibitem[]{Hambly01b} Hambly N.C., Irwin M.J., MacGillivray H.T. 2001b,
MNRAS, 326, 1295.

\bibitem[]{Hambly01c} Hambly N.C., Davenhall A.C., Irwin M.J.,
MacGillivray H.T., 2001c, MNRAS, 326, 1315.

\bibitem[]{Hoffman01} Hoffman Y., Eldar A., Zaroubi S., Dekel A. 2001,
astro-ph/0102190.

\bibitem[]{Jones04} Jones D.H., Saunders W., Colless M., Read M.A.,
Parker Q.A. \etal 2004, MNRAS, 355, 747.

\bibitem[]{Kaldare03} Kaldare R., Colless M., Raychaudhury S., Peterson B.A.
2003, MNRAS, 339, 652.

\bibitem[]{Kraan96} Kraan-Korteweg R.C., Woudt P.A., Cayatte V., Fairall A.P.,
Balkowski C. \etal 1996, Nature, 379, 519.

\bibitem[]{Kurtz98} Kurtz M.J., Mink D.J. 1998, PASP, 110, 943.

\bibitem[]{Lund86} Lund, G. 1986, Optopus. Multiple object spectroscopy at the
Cassegrain focus of the 3.6 M telescope, in ESO Operating Manual, Garching:
European Southern Observatory (ESO)

\bibitem[]{Metcalfe91} Metcalfe N., Shanks T., Fong R., Jones L.R. 1991,
MNRAS, 249, 498.

\bibitem[]{Norberg02} Norberg P. et al. 2002, MNRAS, 336, 907.

\bibitem[]{Oguri04} Oguri M., Takahashi K., Ichiki K., Ohno H. 2004,
astro-ph/0410145.

\bibitem[]{Parker95} Parker Q.A., Watson F.G. 1995, in ``Wide Field
Spectroscopy and the Distant Universe'', 35th Herstmonceux Conference,
ed. S.J. Maddox, \& A. Aragon-Salamanca, (Singapore: World Scientific), 33.

\bibitem[]{Pierre94} Pierre M., B\"ohringer H., Ebeling H., Voges W.,
Schuecker P. \etal 1994, A\&A 290, 725.

\bibitem[]{Quintana95} Quintana H., Ram{\'\i}rez A., Melnick J.,
Raychaudhury S., Slezak E. 1995, AJ 110, 463.

\bibitem[]{Quintana97} Quintana H., Melnick J., Proust D., Infante L.
1997, A\&AS 125, 247.

\bibitem[]{Quintana00} Quintana H., Carrasco E.R., Reisenegger A. 2000, AJ,
120, 511.

\bibitem[]{Quintana05} Quintana H., Proust D., Carrasco E.R., Reisenegger
A. 2005, in preparation.

\bibitem[]{Rag05} Ragone C.J., Merch\'an M., Muriel H., Proust D.,
Reisenegger A., Quintana H. 2005, A\&A, submitted.

\bibitem[]{Ray89} Raychaudhury S. 1989, Nature, 342, 251.

\bibitem[]{Reg89} Reg\"os, E., \& Geller, M. 1989, AJ, 98, 755.

\bibitem[]{Reise00} Reisenegger A., Quintana H., Carrasco E.R., Maze J. 2000,
AJ, 120, 523.

\bibitem[]{Scara89} Scaramella R., Baiesi-Pillastrini G, Chincarini G.,
Vettolani G., Zamorani G. 1989, Nature 338, 562.

\bibitem[]{Stein96} Stein, P. 1996, A\&AS, 116, 203

\bibitem[]{Smoot92} Smoot G.F., Bennett C.L., Kogut A., Wright E.L., Aymon J.
\etal 1992, ApJ, 396, L1.

\bibitem[]{Tody93} Tody D. 1993, in Astronomical Data Analysis Software and
Systems II, A.S.P. Conference Ser., Vol 52, eds. R.J. Hanisch,
R.J.V. Brissenden, \& J. Barnes, 173.

\bibitem[]{Yos01} Yoshida N., Sheth R., Diaferio A. 2001, MNRAS, 328, 669.

\end{thebibliography}
\end{document}